# Frequency Structure of Heart Rate Variability

© Valery Mukhin

*Institute of Experimental Medicine, St. Petersburg, Russia*

*e-mail: Valery.Mukhin@gmail.com*

*Factor structure of heart rate periodogram has been detected with factor analysis. The results showed that there are at least four periodical phenomena of HRV. Two of them have not been discovered and physiologically explained yet. Their frequency ranges are 0.21 to 0.31 1/beat with the peak at 0.26 1/beat and 0.25 to 0.5 1/beat with the peak 0.35 1/beat. Despite of differences of the peak frequencies the frequency rages of the factors are overlapped. Therefore, power of spectral density within any frequency range could not be a measure of a modulating physiological mechanism activity.*

It has known that there is association between heart rate variability and physiological state. This makes it possible to hope to use the analysis of heart rate variability (HRV) as a new quantitative diagnostic means. But, there are not reliable diagnostic techniques based on the HRV analysis so far (Paraty at al., 2006). Probably, "complex and largely undiscovered physiology" (Taylor & Studinger, 2006) of HRV is a cause of this. Physiological validity of the most HRV parameters is doubtful. Thus, there is not sufficient evidence (Vetter, 1998) that there are separate frequency ranges and strong boundaries between them at heart rate periodogram. Authors define them voluntary as most probable frequencies of the two well known heart rate modulation phenomena (respiratory arrhythmia and low frequency oscillations associated with the Traube-Hering-Mayer waves of blood pressure) (table 1). In other words, frequency structure of HRV has not been defined properly.

**Table 1.** Frequency ranges of heart rate periodogram by different authors.

| № | Name | Frequency | Corrected frequency*, $beat^{-1}$ | Author |
|---|---|---|---|---|
| 1 | Respiratory effects | 0.25-0.40 $beat^{-1}$ | -- | Sayers B McA, 1973 |
| 2 | Respiratory frequencies | 0.2-0.3 Hz | 0.13-0.3 | Taylor JA et al., 1998 |
| 3 | Respiratory waves | 0.23-0.19 Hz | 0.12-0.23 | Aksionov VV, 1986 |
| 4 | HI-FR | 0.224-0.280 Hz | 0.149-0.280 | Pomeranz B. at al., 1985 |
| 5 | Mid-frequency | 0.12 Hz | 0.08-0.12 | Akselrod S. at al., 1981 |
| 6 | Low frequency | 0.1 Hz | 0.07-0.10 | Sayers B McA, 1973 |
| 7 | Slow waves 1 | 0.09-0.14 Hz | 0.06-0.14 | Aksionov VV, 1986 |
| 8 | Low frequency | 0.05-0.15 Hz | 0.03-0.15 | Taylor JA et al., 1998 |
| 9 | LO-FR | 0.04-0.12 Hz | 0.026-0.120 | Pomeranz B. at al., 1985 |
| 10 | Low frequency | 0.04 Hz | 0.026-0.04 | Akselrod S. at al., 1981 |
| 11 | Slow waves 2 | 0.027-0.074 Hz | 0.018-0.074 | Aksionov VV, 1986 |
| 12 | Very low frequency | 0.003-0.03 Hz | 0.002-0.03 | Taylor JA et al., 1998 |

\* Corrected (by Author) frequency is a frequency measured in 1/beat units calculated on the base of frequency measured in hertz. Its low bound is equal to initial frequency (Hz) because low bound of normal average heart rate equals to 1 Hz, its upper frequency is initial frequency multiply by 1.5 because the upper bound of normal average heart rate equals to 1.5 Hz.

Actual frequency structure of heart rate variability may be more complex than it is usually considered. Firstly, frequency ranges may overlap. Secondly, there may be any other heart rate modulation mechanisms except of the two well known ones causing latent heart arte modulations within the same frequency ranges or at other ones.

Certainly, the best way to define frequency structure of HRV is physiological investigation of heart rate variability nature. But it may be another way. Exploratory data analysis may make it possible to define factor structure of HRV, and to formulate hypothesis on latent heart rate modulation phenomena.

The aim of this study was to detect the frequency structure of periodical modulations of heart rate in short records.

**Methods.**

ECG in 12 standard leads was recorded in three groups of 64, 39 and 19 healthy volunteers. Short tachograms of 300 cardiointervals were defined.

In order to avoid a contradiction when series of time intervals (RR) are analyzed as a function of the same time we analyze them as a function of number. Therefore, at the abscissa of the spectrogram we had not Hertz bat the reciprocal number of period in cardiocycles.

150 harmonics (N/2=300 cardiointervals/2) were defined with discrete Fourier transform technique. In order to normalize the distribution of the periodogram values at each frequency they were logarithmically transformed. Then factor analysis of the 150 periodogram frequencies was carried out. Factors were extracted by the principal components method with the varimax rotation.

Identity of the factors from different groups was confirmed with the Pearson product-moment correlations.

**Results and discussion.**

The results in each group were similar. Forms of the diagrams of factor loadings of the four first factors are like wave (fig. 1).

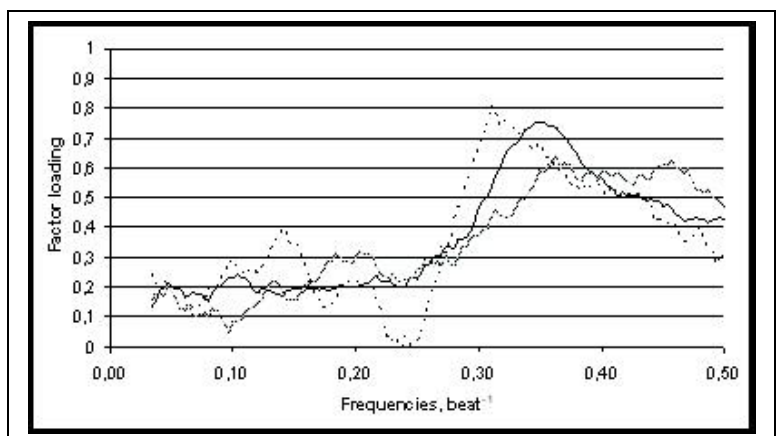

THE FIRST FACTORS

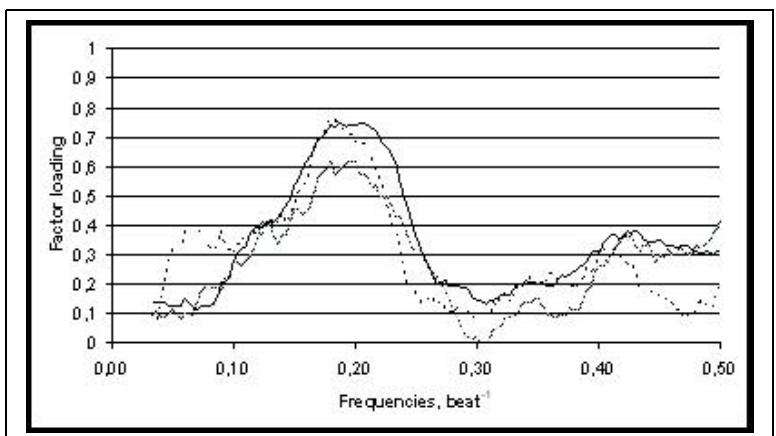
THE SECOND FACTORS

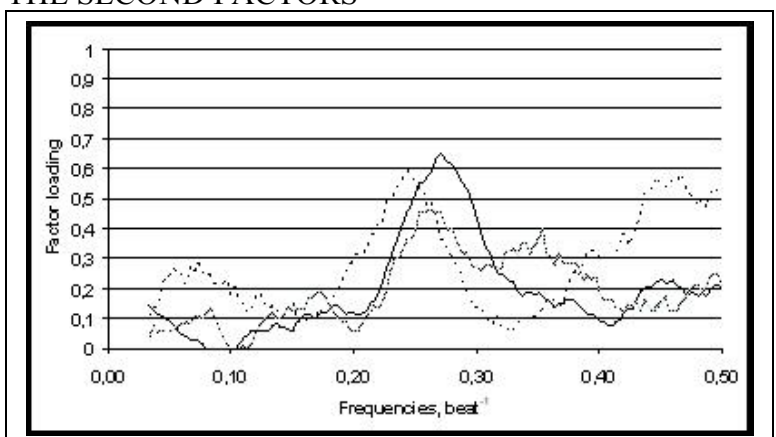
THE THIRD FACTORS

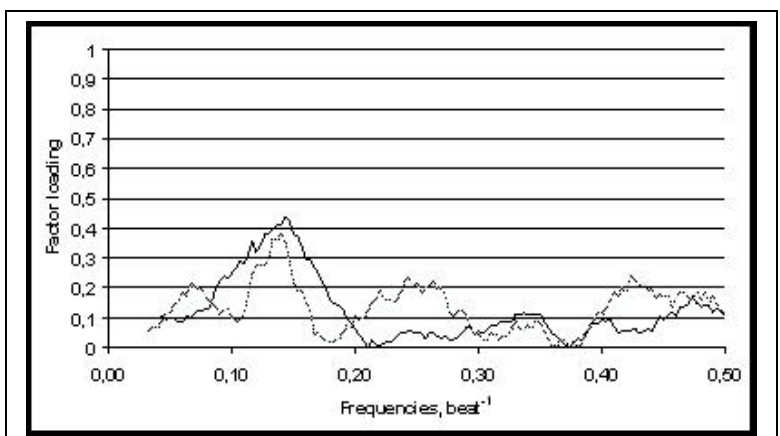
THE FOURTH FACTORS

*Fig. 1. Comparative diagrams of the factors.*
Solid line -- the first group
Dashed line -- the second group
Dotted line -- the third group

Each of the factors had one such wave. The waves of the factor loadings in each group lied approximately at the same frequencies. The first factors from 0.27 to 0.50 1/beat. The second factors from 0.09 to 0.28 1/beat. The third factors from 0.22 to 0.27 1/beat. And the fourth factors from 0.09 to 0.13 beats$^{-1}$. The peaks of the similar waves in different groups were also approximately at the same frequencies. The first factors had the peaks around 0.34 1/beat. The second factors – around 0.19 1/beat. The third factors around 0.26 1/beat. And fourth factors around 0.13 1/beat. Are they the same waves at the same factors among the groups? The coinciding of the waves was confirmed by the significant correlations between the same factors among the groups (table 2).

*Table 2. Correlations between the similar factors among the groups.*

| Pairs of the groups | Factor 1 vs. Factor 1 | Factor 2 vs. Factor 2 | Factor 3 vs. Factor 3 | Factor 4 vs. Factor 4 |
|---|---|---|---|---|
| Group 1 vs. group 2 | 0.81 | 0.79 | 0.33 | |
| Group 1 vs. group 3 | 0.86 | 0.90 | 0.73 | 0.43 |
| Group 2 vs. group 3 | 0.62 | 0.71 | 0.26 | |

*Only correlations with $p<0.05$ is shown.*

If we interpret the waves as physiological phenomena of periodical heart rate modulation, we conclude that there are at least four such phenomena instead of the two ones mainly being discussed now. The fourth factor wave is about from 0.09 to 0.15 1/beat and has peak at 0.13 1/beat. These frequencies coincide with the so called low frequency modulations related with Mayer waves of blood pressure (Hyndman, 1998). The second factor wave is about 0.14 – 0.24 1/beat and peak at 0.18 1/beat. It may only be the frequencies of respiratory sinus arrhythmia. The wave of the third factor is about 0.21 – 0.31 1/beat with peak is at 0.26 1/beat. First factor has the wave at frequencies from 0.25 up to 0.5 1/beat and peak 0.35 1/beat. Last two factors have not physiological interpretation.

Each wave of factor loadings is caused by the certain oscillation phenomena of heart rate. How could we prove it? We can see the real waves in the spectra at the frequencies of the factor loading waves (fig.2).

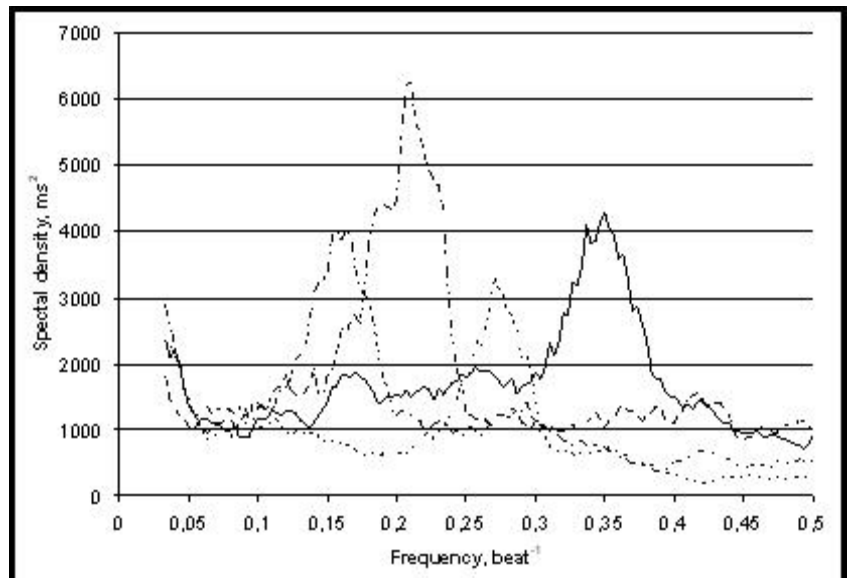

*Figure 2.* The examples of spectra of heart rate variability.
Solid line -- the peak at the frequencies of the first factor.
Dashed line -- the peak at the frequencies of the second factor.
Dotted line -- the peak at the frequencies of the third factor.
Dash-and-dot line -- the peak at the frequencies of the fourth factor.

The existence this oscillation can be confirmed immediately by the example or heart rate where this modulation is well expressed (fig. 3).

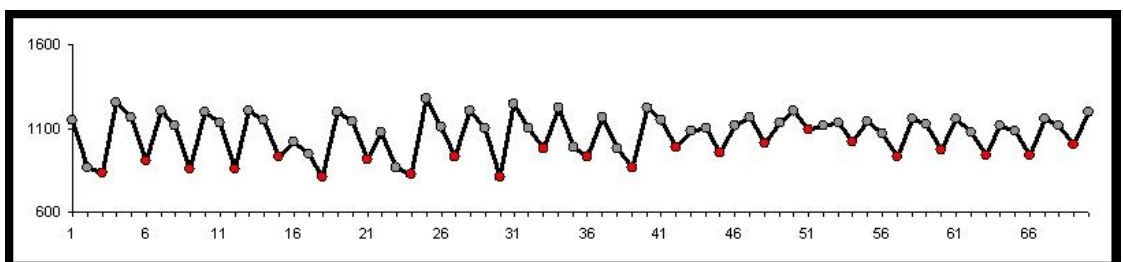

*Figure 3.* A heart rate tachogram. The modulation at 3 beats is very well expressed.

The factor analysis all groups together (N=122) make it possible to have more reliable frequency parameters of the factor loading waves (fig. 4).

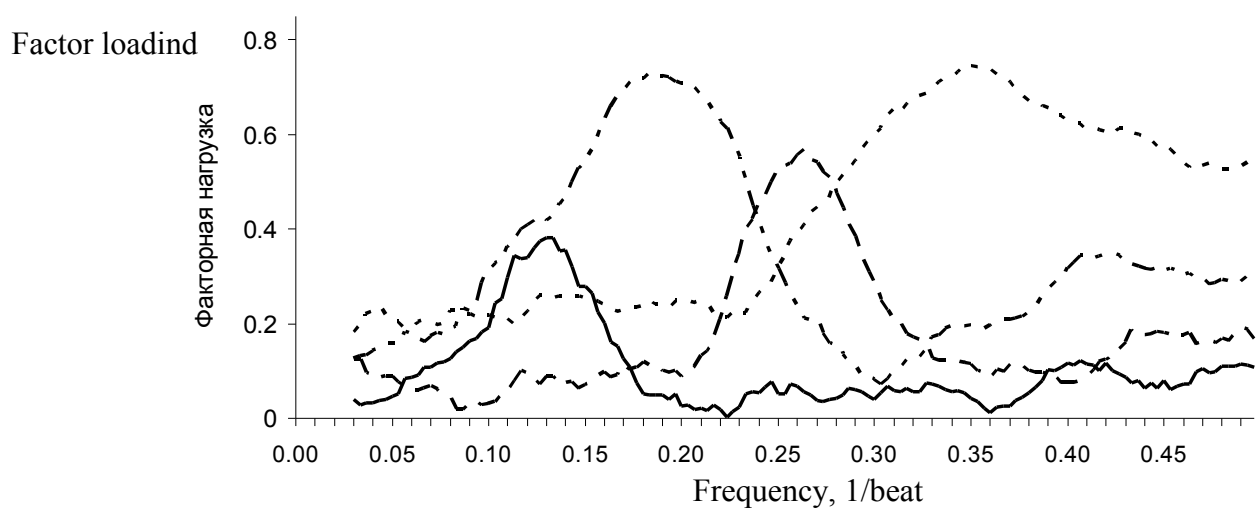

*Figure 4.* Diagrams of the factor loadings. All groups together.

**In conclusion.**

Frequency structure of heart rate variability is more complex than it is considered now. The results make us suppose that there are at least four periodical phenomena of HRV. Two of them have not been discovered and physiologically explained yet. Despite of difference of the peak frequencies the frequency ranges of the heart rate modulation activities is overlapped. Therefore, power of spectral density within any frequency range could not be a measure of a modulating physiological mechanism activity.

**References.**